**A global consumer-led strategy to tackle climate change**

Anthony J. Webster

**A successful response to climate change needs vast investments in low-carbon research, energy, and sustainable development. Governments can drive research, provide environmental regulation, and accelerate global development, but the necessary low-carbon investments of 2-3% GDP have yet to materialise. A new strategy to tackle climate change through consumer and government action is outlined. It relies on ethical investments for sustainable development and low-carbon energy, and a voluntarily financed low-carbon fund for adaptation to climate change. Together these enable a global response through individual actions and investments. With OECD savings exceeding 5% of disposable household income, ethical savings alone have considerable potential.**

Hurricanes, floods, landslides, and droughts are a regular reminder of the risks of climate change (www.reliefweb.int), but a successful response to climate change is inextricably linked to sustainable development, and improvements in health [1,2]. In fact many health professionals regard a reduction in fossil-fuel use as the greatest public health opportunity of this century [1].

A rapid transition from fossil-fuels to low-carbon technologies will cost up to 3% of world GDP [3], and ongoing investments are needed to combat the long-term consequences of climate change [4,5]. Government or industry can raise this through an energy tax or levy [4,5], but this has yet to happen. It seems prudent to consider alternative approaches.

**Ethical investments for sustainable development and a low-carbon transition**

Ethical consumer savings and investments are comparatively new products. They vary between low-carbon investment funds that aim to provide high returns, and development funds that solely aim to return investments as savings. Both rely on invested capital, but the latter are motivated more by philanthropy than personal gain. Ethical savings have considerable potential. As an average over the OECD countries, savings presently account for over 5% of disposable household income [6].

Development funds such as Shared Interest offer support through cheap loans, as opposed to a "hand-out" in aid, while offering savers the prospect of getting their money back. International development is not prioritised by climate funds, but is essential for decent living standards and sustainable populations. Ethical development funds offer a market-led response to poverty, complementary to the traditional governmental and charity-led approaches.

International development seems integral to a successful long-term strategy. Whereas global population growth can derail efforts at adaptation and mitigation, population sizes are expected to stabilise if the Sustainable Development Goals are met [7,8]. Adaptation efforts are also hampered by poverty, with many disasters caused by settlement in inappropriate locations [9-11]. Inadequate planning regulations or building standards are partly to blame [9], but habitation in vulnerable areas is often driven by need. This is true in all countries, but more so in developing ones [10,11].

Low-carbon investment must be made easier. Existing products can be confusingly diverse, and difficult for consumers to invest in. New, simplified products are needed, and offer huge potential for growth. They must be popularised and encouraged. Global icons and governments can help through visible support and legislation to encourage greater low-carbon and ethical investment.

**Voluntary donations for adaptation to climate change**

Personal investments will accelerate a low-carbon transition, but are unlikely to provide revenues for adaptation to climate change. Long-term revenues for adaptation are needed in the post fossil-fuel era. One option is a voluntarily-paid levy that is invested in stocks and other forms of low-carbon technology ownership. It would be led by multi-national brands and high-profile individuals. An invested levy gives twice – it stimulates the area in which it is invested, and returns dividends to pay for long-term adaptation [5]. The levy can be collected by a bank, insurance company, investment fund, or an existing climate fund. It can start arbitrarily small, and grow arbitrarily large.

Present climate funds are financed by government donations, with a focus on mitigation and adaptation in developing countries (www.climatefundsupdate.org). Most funding is allocated as grants or loans, and offer limited return on investments. Government donations need the support of electorates, whose attitudes to charity and climate change vary considerably. The result is concern about a lack of "replenishment policies" for longer-term funding. Research and development is overlooked, as is the need for a low-carbon transition in developed countries.

Contributions must be easy to make. Small regular contributions that are easy to absorb into everyday living expenses seem preferable, like regularly adding a small tip to a bill. A direct levy on all purchases, possibly through government taxation is one option, but a levy could be collected through the distributor. For example, Amazon Smile donates 0.5% of purchase price to a consumer's chosen charity, and Ethical Superstore allows you to add a donation at checkout. Dedicated credit cards that automatically allocate a portion of expenditure to the levy are another option.

Indirect payments can be made by purchasing suitably branded products, whose producers pass on the additional costs to consumers. Labelling analogous to a Fairtrade mark can indicate participating products, encouraging consumers to select them. Participation could involve an entire brand or a new, differentiated product, and can include insurance, energy, drinks, and clothing. Price increases of around 1% to 2% would be difficult to distinguish from natural price variations. If paid by the largest and most profitable multinational companies it would raise a considerable amount.

There is a huge marketing opportunity for the global brands who lead in paying a voluntary levy. An overwhelming global demonstration of public support would incentivise other companies to follow. Multinational brands such as Coca Cola, Google, and Microsoft are ideal candidates, as are credit card companies, energy companies, insurance companies, or governments. The best ways of collecting a voluntarily paid levy must be explored, and a climate- or investment-fund must take the lead in co-ordinating investments to ensure long-term revenues for adaption.

**The advantages of a consumer-driven response**

There are advantages to consumer-led initiatives. Voluntary financing avoids the need for complex agreements between governments with different priorities. Different countries have different resources, different challenges, and as a result, different priorities. Personal responses to climate change depend on aversion to risk and compassionate responsibility for others, future and present. Our views will differ. Voluntary donations and investments help overcome these differences - only those that choose to pay, need to pay.

Poorer countries cannot be expected to sacrifice development for long-term environmental concerns. Similarly, it is unrealistic to expect people to reduce their living standards without recompense. Voluntary payments need no additional laws to be passed, and contributions will be from those who can afford them. Global economies will be stimulated by low carbon investments and protected by adaptation measures such as improved flood defences, benefiting everyone.

The growing low-carbon investments are likely to increase the value of existing stock and incentivise further private investment. Ongoing legal threats and falling competitiveness will reduce confidence in fossil-fuel investments [5,12]. Together this will accelerate the shift to low-carbon technologies and amplify consumer investments well beyond their original value.

**Governments must drive research and ensure developments are sustainable**

Consumer action can increase finance for low-carbon investments, but greater research and development is needed. There are insufficient good low-carbon investment opportunities, limiting large-scale investments [13]. Governments can help. Developments can improve the efficiency of energy production, storage, conversion, and use, including industrial applications such as chemical engineering [14,15]. If energy is cheap enough then fossil-fuel extraction will reduce. If low-carbon investments are plentiful and lucrative enough then divestment from fossil-fuel stocks will increase [13]. Low-carbon research must expand, and be allowed to be more innovative and ambitious.

Governments must encourage new technologies that can use intermittent excesses of e.g. solar energy. Energy can be stored in many forms – heat, cold, chemical, gravitational, but it may be possible to do better. Technologies must be developed that use energy surpluses to extract carbon dioxide from emissions or the atmosphere for use in agriculture or chemical production [14,15].

Geological scale coverage is needed to capture enough of the sun's energy, either directly as light, or indirectly through wind, wave, and precipitation [16]. This requires sustainable practices for farming, industry, and energy production. Governments must ensure these vast, widely distributed activities are beneficial for the climate and the wildlife that depend on it. Long term success needs high calibre research, driven by a combination of government-led, business-led, and consumer-led initiatives.

**Future priorities**

Ethical savings and a voluntary levy are both scalable, and must be popularised. Ethical savings can redirect a substantial part of the 5% OECD disposable household income that is saved, into low-carbon and development-related investments. A low carbon levy might regularly collect 1-2% of transactions in donations to pay for long-term adaptation needs. Multinational brands are needed to lead the collection of a levy, and a suitable climate or investment fund is needed to invest it.

Global development must be prioritised by governments, as must research and development. Private-sector research is likely to grow, but cannot be relied on. Innovations in low-carbon energy offer the greatest potential for a rapid and successful transition.

Climate change requires global action. Ethical investment and adaptation funds can provide consumers with the power to drive change. If climate change is tackled through global development and a low-carbon transition, it can be the catalyst for an equitable and sustainable future.

[1] N. Watts et al. Lancet, **386**: 1861–1914 (2015).

[2] M. Springmann, H.C.J. Godfray, M. Rayner, P. Scarborough, PNAS, **113**, no. 15, 4146-4151, (2016).

[3] N. Fabian, Nature, **519**, 27-29, (2015).

[4] R.H. Clarke, "Predicting the price of carbon", (Predict Ability, 2016).

[5] A.J. Webster, R.H. Clarke, Nature, 549, 152-154, (2017).

[6] http://data.oecd.org/hha/household-savings.htm

[7] G.J. Abel, B. Barakat, K.C. Samir, and W. Lutz, PNAS, **113**, No. 50, 14294-14299, (2016).

[8] W. Lutz, R. Muttarak, E. Striessnig, Science, **346**, Iss. 6213, 1061-1062, (2014).

[9] Institute of Mechanical Engineers (IMECHE), "Natural disasters: saving lives today, building resilience for tomorrow", 1 Birdcage walk, Westminster, London SW1H 9JJ, Oct. (2013).

[10] B. Jongman, H.C. Winsemius, J.C.J.H. Aerts, E.C. de Perez, M.K. van Aalst, W. Kron, P.J. Ward, PNAS, **18**, 112, E2271-E2280, (2015).

[11] J. Wu, G. Han, H. Zhou, N. Li, Environ. Res. Lett., **13**, 034013, (2018).

[12] Marjanac, S., Patton, L., & Thornton, J. Nature Geosci, **10**, 616-619, (2017).

[13] M. Golnaraghi, "Climate Change and the Insurance Industry: Taking Action as Risk Managers and Investors", Geneva Association, Talstrasse 70, CH-8001, Zurich, (2018).

[14] P. Lanzafame, S. Abate, C. Ampelli, et al. ChemSusChem, **10**, 4409-4419, (2017).

[15] D.W. Keith, G. Holmes, D. St.Angelo, K. Heidel, Joule, **2**, 1-22, (2018).

[16] D.J.C. MacKay, "Sustainable energy without the hot air", UIT Cambridge, (2008).

**Table 1**: Low-carbon investments, development funds, and voluntary donations for adaptation to climate change, serve complementary purposes.

|  | Drive a low-carbon transition | Accelerate global development | Support long-term adaption |
| --- | --- | --- | --- |
| Low-carbon investments | Yes | Indirectly | No |
| Development funds | No | Yes | Indirectly |
| Voluntary donations (invested in low-carbon technologies) | Yes | Indirectly | Yes |